# High Throughput Parameter Estimation and Uncertainty Analysis Applied to the Production of Mycoprotein from Synthetic Lignocellulosic Hydrolysates

Mason Banks [a], Mark Taylor [b], Miao Guo [a, *]

[a] Department of Engineering, Faculty of Natural Mathematical & Engineering Sciences, King's College London, Strand, London, WC2R 2LS, United Kingdom
[b] Fermentation Lead, Marlow Ingredients, Nelson Ave, Billingham, North Yorkshire, TS23 4HA, United Kingdom

*Corresponding author: miao.guo@kcl.ac.uk

**Abstract**
The current global food system produces substantial waste and carbon emissions while exacerbating the effects of global hunger and protein deficiency. This study aims to address these challenges by exploring the use of lignocellulosic agricultural residues as feedstocks for microbial protein fermentation, focusing on *Fusarium venenatum* A3/5, a mycelial strain known for its high protein yield and quality. We propose a high throughput microlitre batch fermentation system paired with analytical chemistry to generate time-series data of microbial growth and substrate utilisation. An unstructured biokinetic model was developed using a bootstrap sampling approach to quantify uncertainty in the parameter estimates. The model was validated against an independent dataset of a different glucose-xylose composition to assess the predictive performance. Our results indicate a robust model fit with high coefficients of determination and low root mean squared errors for biomass, glucose, and xylose concentrations. Estimated parameter values provided insights into the resource utilisation strategies of *Fusarium venenatum* A3/5 in mixed substrate cultures, aligning well with previous research findings. Significant correlations between estimated parameters were observed, highlighting challenges in parameter identifiability. This work provides a foundational model for optimising the production of microbial protein from lignocellulosic waste, contributing to a more sustainable global food system.

1. Introduction

The current global food system produces substantial waste and carbon emissions and relies on excessive use of arable land and freshwater supplies (Holden et al., 2018). These factors not only result in environmental degradation but also exacerbate the issues of increasing global hunger and protein deficiency according to a recent meta-analysis by Van Dijk et al. (2021). A potential solution to this challenge was explored by Durkin et al. (2022) who demonstrated high potential for carbon recovery from abundant global waste streams such as lignocellulosic agricultural residues, by using them as feedstocks for microbial protein fermentation to produce sustainable and high-quality food-grade protein.

One candidate strain for such a system is *Fusarium venenatum* A3/5 (Fig. 1), a filamentous fungal species notable for its fibrous, meat-like texture and high nutritional value, containing 45-54% protein on a dry basis and all essential amino acids (Coelho et al., 2020; Wiebe, 2002).

These factors in addition to rapid growth rates have given this strain large-scale commercial utility as a microbial protein food product, sold as Quorn mycoprotein (Marlow Ingredients Ltd.), Mycofood (Eternal), and whole-cut mycoprotein by MyForest Foods. (Upcraft et al., 2021) demonstrated the technoeconomic viability and environmental benefits of a large-scale bioprocess system for the production mycoprotein from rice straw, a prominent agricultural residue rich in lignocellulosic biomass.

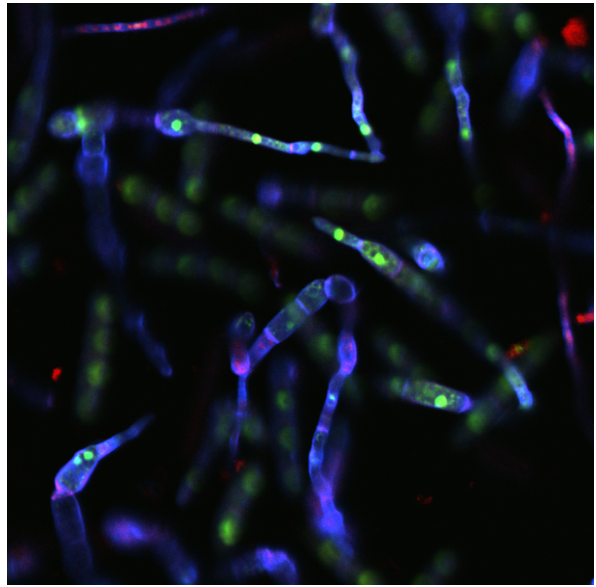

**Fig. 1.** Germinating conidia of Fusarium venenatum A3/5 wild-type (WT) strain. Cell walls (blue), nucleic acids (green), and extracellular proteins (red) were stained with epifluorescent dyes Calcofluor White Stain, SYTO-9 Green Fluorescent Nucleic Acid Stain, and FilmTracer SYPRO Ruby Biofilm Matrix Stain respectively. Image captured using a Nikon AXR w/NSPARC confocal laser scanning microscope, integrated with a Ti2-E inverted microscope equipped with a 20X objective lens.

However, despite increasing research interest in lignocellulosic waste carbon valorisation using microbial protein technologies, several critical bottlenecks remain during bioprocess development that hinder rapid and viable scale-up (Banks et al., 2022; Piercy et al., 2023). One such bottleneck is the development of accurate kinetic models to reliably predict performance dynamics in highly complex systems (Narayanan et al., 2019). The composition of agricultural residues is dependent on plant species and processing techniques, with hydrolysate feedstocks containing a diverse array of substrates and inhibitory compounds, necessitating the use of highly non-linear models with many parameters to be estimated simultaneously (Panikov, 2021). This effectively exacerbates the outstanding challenges related to parameter identifiability and model generalisability (Wieland et al., 2021).

Given the critical global challenges associated with food security and environmental sustainability, and the bioprocess design bottlenecks identified, the research aims of this study are to develop a predictive model for production of *Fusarium venenatum* A3/5 utilising a synthetic mixture of two predominant lignocellulosic substrates, glucose and xylose. We propose a high throughput modelling framework, combining microlitre batch fermentation experiments with analytical chemistry techniques for data collection, while advanced optimisation methods and bootstrap sampling are used to estimate parameters and their uncertainties. Specifically, the current study presents new research addressing following objectives:

1. To assess the effectiveness and reproducibility of a high throughput microlitre batch fermentation system paired with advanced analytical chemistry (UHPLC) for generating time-series data of microbial growth and substrate utilisation. This will involve detailed analysis of the variability in experimental results across different growth phases, providing insights into the robustness of the method.
2. To establish a biokinetic model for *F. venenatum* A3/5 that accurately describes its growth kinetics on mixed substrates emulating lignocellulosic hydrolysates. This includes parameter estimation using a bootstrap approach to quantify uncertainty and ensure robust model predictions.
3. To perform a comprehensive uncertainty and correlation analysis of the estimated parameters. This will help in understanding the interdependencies among parameters and their impact on model reliability and predictive performance, addressing key challenges in parameter identifiability and uniqueness.
4. To validate the generalisability and predictive performance of the developed biokinetic model using an independent dataset from a different glucose-xylose ratio. This step is crucial to demonstrate the model's applicability across varying lignocellulosic composition and its potential utility in optimising microbial protein production processes.

By achieving these aims, this research seeks to provide a good foundation for scalable production of microbial protein from lignocellulosic waste resources, contributing to a more sustainable and efficient global food system.

## 2. Materials and Methods

### 2.1. Experimental methodology and design

Fig. 2 provides an overview of the methodological workflow described in detail in the following subsections. Sterile technique was employed throughout all experiments and performed in a class II biological safety cabinet (ESCO) where appropriate.

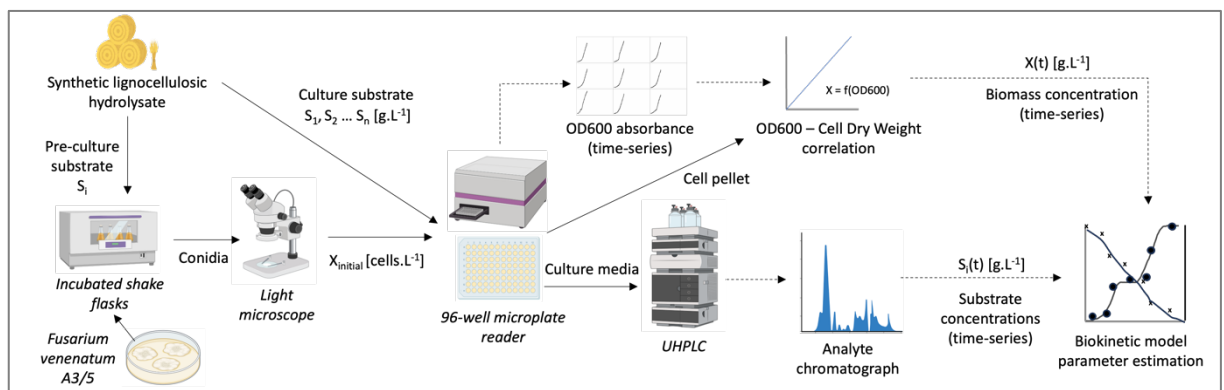

**Fig. 2.** Overview of the experimental workflow for high throughput data generation used to estimate the biokinetic model parameters for *F. venenatum* A3/5 growth from synthetic lignocellulosic hydrolysates. Solid arrows represent material flows and dashed arrows indicate data flows. Image was created using BioRender software.

### 2.1.1. Media preparation

Two sugar stock solutions were prepared by dissolving D-(+)-glucose and D-(+)-xylose (≥99%, Sigma Aldrich) in 1L ultrapure water (Arium Pro, Sartorius, Germany) to a final concentration of 30 w/v %. Nutrient media stock was prepared, and pH adjusted to 6.0 using 50 w/v% sodium hydroxide solution (VWR). Stock solutions were autoclaved for 15 minutes at 121 °C and checked regularly for signs of contamination.

### 2.1.2. Preculture of Fusarium venenatum A3/5

*Fusarium venenatum* A3/5 wild-type (WT) strain was stored medium-term as malt extract agar plate inoculums and refrigerated at 4 °C. A sterile loop was used to inoculate a 50 mL Erlenmeyer flask containing 10 mL of nutrient media with glucose at a concentration of 3 w/v %. The flask was sealed with sterile cotton wool and placed in a shaking incubator (ES-20, Grant-Bio, UK) at 28 °C and 130 rpm for 96 hours. The flask contents were subsequently filtered using a 100 µm cell strainer to obtain a conidia suspension. A Neubauer counting chamber (Hawksley) was used to determine the conidia concentration in the filtrate, which was subsequently diluted using sterile nutrient media to obtain a stock concentration of $10^5$ conidia / mL. The dilute stock was used for short-term storage of *F. venenatum* A3/5 and refrigerated at 4 °C.

### 2.1.3. Batch microlitre fermentation of Fusarium venenatum A3/5

Aliquots from the sugar stock solutions were combined to achieve the desired glucose to xylose ratios of 2:1 and 1:1. 15 µL of both the sugar mixture and the conidia stock solution were added to the study wells of a transparent, sterile, flat-bottom, untreated 96-well plate (VWR) and diluted with 120 µL sterile nutrient media. The total working volume of each well was 150 µL with initial concentrations of $10^4$ conidia / mL and 3 w/v % total sugars.

Outer wells were filled with 150 µL sterile nutrient media as blanks, and triplicate wells of inoculated nutrient media containing no sugar served as negative controls. The lid of the 96-well plate was attached and sealed around the edges using parafilm to minimise evaporation. The plate was then placed into the tray of a microplate reader (FLUOstar Omega, BMG Labtech, Germany) and incubated in stationary mode at 28 °C. Optical density at 600 nm (OD600) was recorded from direct optic bottom absorbance readings taken at 20-minute intervals, providing an indirect measurement of the biomass growth in each well.

### 2.1.4. Time-series sampling protocol and experimental design

The experimental design included 17 time points over 122 fermentation hours, with each time point assigned triplicate well cultures for both 2:1 and 1:1 glucose-xylose conditions, resulting in 102 samples (3 replicates × 17 time points × 2 conditions) across two plates.

At each time point, the 96-well plate was removed from the microplate reader. The sample contents were transferred to sterile 1.5 mL tubes and centrifuged for 5 minutes at 1200 rpm. The supernatant was filtered through 0.2 µm regenerated cellulose filters (Sartorius) directly into 0.3 mL short thread ND9 borosilicate glass vials (VWR) for sugar analysis via ultra high-performance liquid chromatography (UHPLC) (Nexera X3, Shimadzu, Japan).

For UHPLC analysis, 10 µL of each sample was injected into an Aminex HPX-87H carbohydrate analysis column (Bio-Rad) with a 0.005 M sulfuric acid mobile phase, a flow rate of 0.6 mL/min, and a column oven temperature of 60°C. A refractive index detector (RID-20A, Shimadzu, Japan) quantified the eluted glucose and xylose at 30°C.

Control analyses were performed on ultrapure water and blank nutrient media using the same protocol, and substrate concentration data was quantified using 5-point external calibration curves using reference standards for D-(+)-glucose and D-(+)-xylose (Monosaccharides Kit, Sigma Aldrich) across the range of study concentrations. The cell pellets obtained during centrifugation were used to generate an OD600 to cell dry weight correlation (Section 2.1.5). Raw substrate concentration and OD600 data for the fermentation time series under both experimental conditions (1:1 and 2:1 glucose-xylose) are included in Supplementary Materials.

### 2.1.5. $OD600$ to cell dry weight ($CDW$) correlation

To quantify the biomass concentration in the fermentation experiments, a correlation between OD600 and cell dry weight (CDW) was established. This correlation allows for the conversion of OD600 measurements, commonly used in microbial growth studies, into biomass concentrations. For each time point, cell pellets from replicate wells were pooled to ensure a representative sample. The pooled cell pellets were transferred onto pre-dried, pre-weighed grade 698 glass fibre filters (VWR). The cell pellets were rinsed ten times with ultrapure water under vacuum filtration to remove any residual media components. The filters containing the cell pellets were oven dried at 80°C for 24 hours. After drying, the filters were weighed using an analytical balance (Quintix Pro, Sartorius, Germany) to obtain the dry weight of biomass. The recorded weight was normalised by the initial culture volume to obtain the cell dry weight (g/L). This procedure was repeated across all time points to gather sufficient data for correlation analysis. The corresponding OD600 values for each time point were obtained by averaging the measurements from the replicate wells. Regression analysis was performed to derive the correlation equation between CDW and OD600, presented by Equation (1):

$$\text{CDW} = 0.2927 \cdot \text{OD600}^3 + 0.2631 \cdot \text{OD600}^2 + 1.2808 \cdot \text{OD600} - 0.1596 \qquad (1)$$

The cubic relationship between CDW and OD600 was selected due to the high coefficient of determination ($R^2$ = 0.9603) when compared with linear and quadratic forms. This correlation was used to convert OD600 measurements into biomass concentrations, $X$, for the fermentation studies and subsequent parameter estimation. Further details of the analysis are provided in Supplementary Materials.

### 2.2. Parameter estimation and bootstrap uncertainty analysis

The experimental data obtained using the methods detailed in Section 2.1 were utilised for the subsequent parameter estimation and bootstrap uncertainty analysis to develop a robust and predictive biokinetic model. All analyses were performed using Python (version 3.9.6) with libraries including NumPy, SciPy, and Pandas. The computations were executed using a Dell Precision 7760 laptop equipped with 11th Gen Intel Core i9-11950H CPU @ 2.60GHz, 128GB of RAM, and a NVIDIA RTX A3000 GPU, running on Windows 10.

### 2.2.1. Biokinetic model system formulation

The parameter estimation problem for the mixed substrate fermentation model was formulated using a system of ordinary differential equations (ODEs) representing the dynamic behaviour of biomass ($X$), glucose ($S_1$), and xylose ($S_2$) concentrations. The unstructured kinetic model system was adapted from the dual substrate model derived by Vega-Ramon et al. (2021). The ODEs are given by Equations (2) – (4), and parameters to be estimated are listed and described in Table (1).

$$\frac{dX}{dt} = \left(\frac{\mu_{m1} S_1}{S_1 + K_{c1} X}\right) X + \left(\frac{\mu_{m2} S_2}{S_2 + K_{c2} X} \cdot \frac{1}{1 + k_I S_1}\right) X \tag{2}$$

$$\frac{dS_1}{dt} = -Y_{S1} \left(\frac{\mu_{m1} S_1}{S_1 + K_{c1} X}\right) X \tag{3}$$

$$\frac{dS_2}{dt} = -Y_{S2} \left(\frac{\mu_{m2} S_2}{S_2 + K_{c2} X} \cdot \frac{1}{1 + k_I S_1}\right) X \tag{4}$$

**Table 1**. List of seven parameters to be estimated for the dual substrate fermentation model system with their physical interpretations and units.

| Parameter | Description | Units |
|---|---|---|
| $\mu_{m1}$ | Maximum specific growth rate (glucose) | h$^{-1}$ |
| $K_{c1}$ | Half-saturation constant (glucose) | g·L$^{-1}$ |
| $\mu_{m2}$ | Maximum specific growth rate (xylose) | h$^{-1}$ |
| $K_{c2}$ | Half-saturation constant (xylose) | g·L$^{-1}$ |
| $k_I$ | Inhibition constant (inhibition of xylose uptake by glucose) | g·L$^{-1}$ |
| $Y_{S1}$ | Yield coefficient (glucose) | g (glucose)·g (biomass)$^{-1}$ |
| $Y_{S2}$ | Yield coefficient (xylose) | g (xylose)·g (biomass)$^{-1}$ |

### 2.2.2. Objective function definition

The objective function to be minimised was defined as the sum of mean squared errors (MSE) between the model predictions and experimental data for each of the three state variables, $X$, $S_1$, and $S_2$, with an added L2 regularisation term to mitigate overfitting and penalise extreme solutions. The MSE formula and objective function are given by Equations (5)-(6) respectively.

$$\text{MSE} = \frac{1}{n} \sum_{i=1}^{n} [(X_{exp,i} - X_{pred,i})^2 + (S_{1\,exp,i} - S_{1\,pred,i})^2 + (S_{2\,exp,i} - S_{2\,pred,i})^2] \tag{5}$$

$$\text{Objective Function} = \text{MSE} + \lambda_{reg} \sum_{j=1}^{p} \theta_j^2 \tag{6}$$

where n is the number of time points, $\lambda_{reg}$ is the regularisation strength set to 0.1, and $\theta_j$ are the model parameters.

### 2.2.3. Solving the initial value problem

The model ODEs given by Eq. (2) – (4) were solved numerically using the 'solve_ivp' function (scipy.integrate). This function was used to integrate the system of ODEs over the time series intervals using the initial state variable conditions from the experimental data. The parameters of the model system were passed as arguments to the function, and the integration was performed using the LSODA method (which automatically switches between stiff and non-stiff solvers) with relative and absolute tolerances set to 1E-5 and 1E-8 respectively. The logarithmic transformation of parameters was implemented to ensure positive values during optimisation and to improve stability of the numerical solver.

### 2.2.4. Stochastic global optimisation using differential evolution (DE) algorithm

The parameter estimation problem was solved using the differential evolution (DE) algorithm, specifically the 'differential_evolution' function (scipy.optimise) was used. DE is a stochastic global optimisation algorithm well-suited the highly non-linear and convex regression problems posed by microbial kinetic model systems (Ashraf and Pfaendtner, 2020; Fang et al., 2009; Manheim et al., 2019; Sumata et al., 2013). The DE algorithm was configured with the following hyperparameters: a population size of 50, a mutation factor range of (0.5, 1), and a recombination rate of 0.7. This configuration was selected based on a comparative evaluation of the algorithm performance and robustness with different combinations of hyperparameters (see Supplementary Materials).

### 2.2.5. Biokinetic model calibration, uncertainty and correlation analysis via bootstrap sampling of 1:1 glucose-xylose data

The data sets for the two study conditions, 1:1 and 2:1 glucose-xylose ratios, were used for model calibration and validation respectively. To improve the uniqueness and identifiability of parameter estimates, the data representing the initial lag phase of biomass growth was removed prior to the model calibration step using 1:1 glucose-xylose data. The truncated dataset began at a fermentation time of 49.3 hours. This preprocessing step was necessary due to the significant influence of the extended lag phase on the stability of the optimisation algorithm.

To quantify the uncertainty in the parameter estimates, a bootstrapping approach was employed to repeatedly sample with replacement from the original time series dataset for the 1:1 glucose-xylose condition, thereby creating multiple bootstrap samples. Parameter estimation was performed for each bootstrap sample yielding a distribution of estimates. Given the initial set of experimental data $D$, bootstrapping generated B replicated datasets $D_1^*, D_2^*, \ldots, D_B^*$. For each dataset $D_b^*(b = 1, 2, \ldots, B)$, the DE algorithm estimated the parameter set $\theta_b^*$. A total of 1500 bootstrap samples were taken from the experimental data set using random sampling with replacement. Statistical properties mean, median and confidence intervals of the parameter distributions were then calculated. This method was used to obtain a robust estimation of parameter uncertainties (and the corresponding uncertainty in model outputs) by leveraging the variability identified in the experimental data set.

For the correlation analysis, no assumptions were made regarding the shape of the bootstrap parameter estimate distributions or the type of dependency between parameters

(other than monotonicity). Hence, Spearman's rank correlation coefficient ($\rho$) was used to assess the relationship between pairs of parameters, calculated as:

$$\rho = 1 - \frac{6 \sum d_i^2}{n(n^2 - 1)}$$

where $d_i$ is the difference between the ranks of each pair of values and $n$ is the number of bootstrap samples.

The correlation matrix and corresponding p-values were computed for each pair of parameters, and the significance of the correlations was determined through hypothesis testing at levels 0.05, 0.01, and 0.001. The heatmap of Spearman correlation coefficients with significance annotations was generated to illustrate the strength and direction of the pairwise parameter relationships. Detailed results of the hypothesis tests including p-values are provided in the Supplementary Materials.

### 2.2.6. Biokinetic model validation with 2:1 glucose-xylose data

The generalisability of the parameter estimate distributions and predictive performance of the model were cross-validated using the independent dataset from the 2:1 glucose-xylose fermentation condition. The model was simulated with initial values from the 2:1 data set for each of the 1500 parameter sets obtained from the calibration step. The predicted values were compared to the experimental data from the 2:1 condition. Statistical metrics quantifying goodness of fit were calculated for state variables $X$, $S_1$, and $S_2$ to determine the model's predictive accuracy, namely the coefficient of determination ($R^2$), root mean squared error (RMSE) and mean absolute error ($MAE$) as defined by Equations (7)-(9).

$$R^2 = 1 - \frac{\sum_{i=1}^{n}(y_i - \hat{y}_i)^2}{\sum_{i=1}^{n}(y_i - \bar{y})^2} \tag{7}$$

$$\text{RMSE} = \sqrt{\frac{1}{n}\sum_{i=1}^{n}(y_i - \hat{y}_i)^2} \tag{8}$$

$$MAE = \frac{1}{n}\sum_{i=1}^{n}|y_i - \hat{y}_i| \tag{9}$$

Where for each time point $i$, $y_i$ is the mean of replicate experimental data and $\hat{y}_i$ is the corresponding model prediction. $\bar{y}$ is the overall mean of the observed data, and $n$ is total number of time points.

The validation step was necessarily included to determine the extent of overfitting and to evaluate the model capacity in describing the observed dynamics for different initial substrate concentrations.

## 3. Results and discussion

*3.1. Evaluation of high throughput microlitre batch fermentation*

Figures 3(a)-(c) and 3(d)-(f) show the time-series results of the batch microlitre fermentation for both 1:1 and 2:1 glucose-xylose initial conditions respectively.

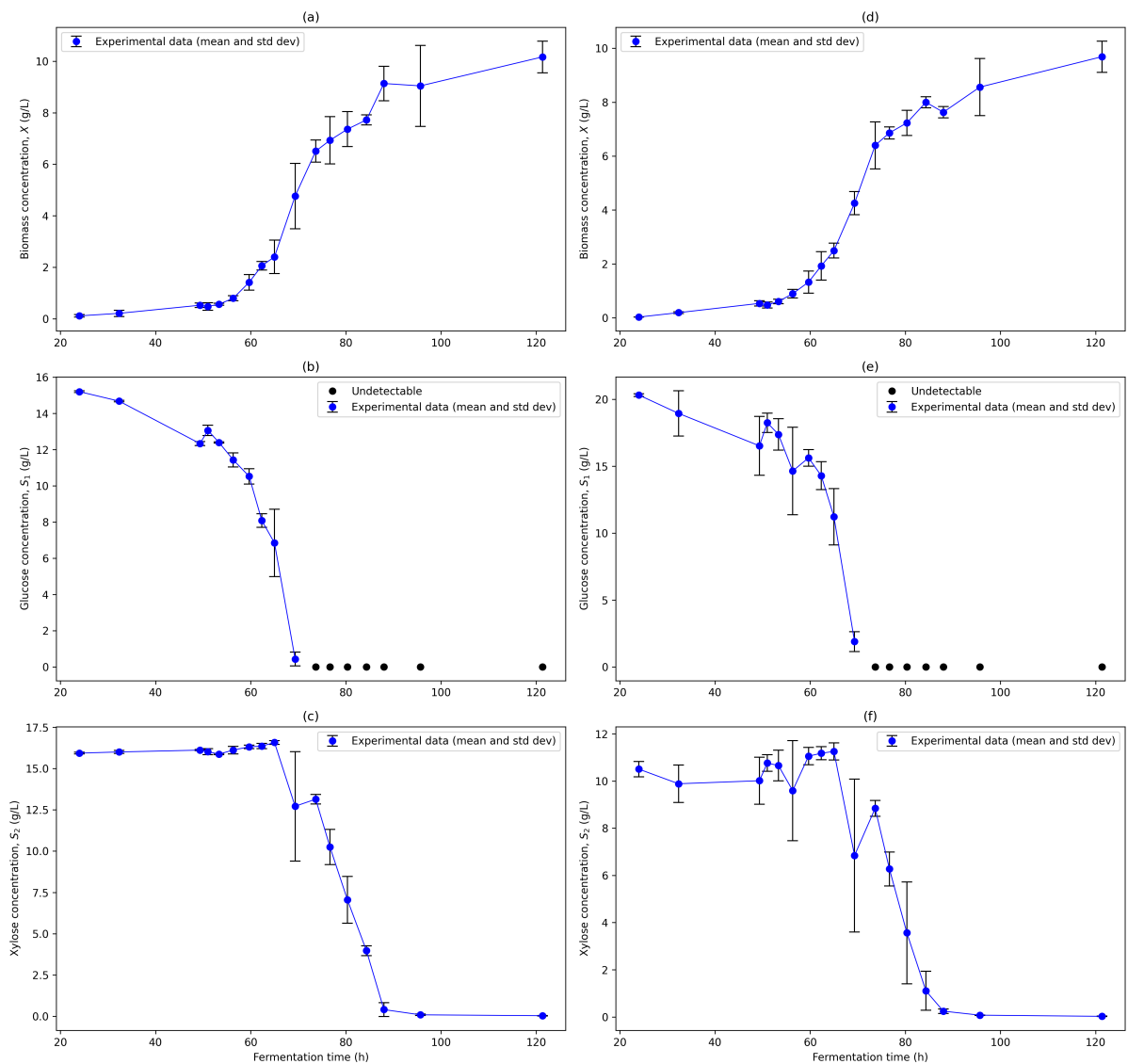

**Fig. 3.** Experimental Data for Mixed Glucose-Xylose Microlitre Batch Fermentations. Experimental data for biomass, glucose and xylose concentrations are shown for 1:1 glucose-xylose (a)-(c) and 2:1 glucose-xylose (d)-(f) initial conditions. Mean (blue dot) and standard deviation (black bars) for are shown for each timepoint. Undetectable glucose concentrations are also shown (black dot).

### 3.1.1. *Resource utilisation and biomass growth of F. venenatum A3/5*

For both conditions there was a period of slow linear increase in biomass coupled with glucose utilisation between 24 and 51 hours, while there was no significant consumption of xylose. The protracted lag phase may be a result of adaptive responses (i.e. signal activation, transcriptomic changes) to the stress incurred by the switch between glucose pre-culture and mixed substrate culture environments (Hamill et al., 2020). Furthermore, media composition and lag phase duration have been demonstrated to influence the time delay of conidia germination and the rate of early hyphal extension in fungal populations (Gougouli and Koutsoumanis, 2013; Meletiadis et al., 2001).

Glucose consumption increases rapidly during the first log phase between 51 to 73.7 h, during which the specific growth of biomass reaches its fastest rate across the entire fermentation time course. The transition between the first and second log phase is indicated by the complete exhaustion of glucose and an abrupt decrease in the biomass growth rate between 69.3 and 73.7 h. Interestingly, there is no obvious second lag phase of arrested growth typically associated with a diauxic shift. Instead, the biomass enters a second log phase of slower growth (73.7 to 88 h) where xylose is rapidly utilised.

However, the data for both conditions indicate that xylose consumption begins earlier during the first exponential growth phase (65 to 69.3 h), although the extent to which is unclear due to the large variability in replicate $S_2$ measurements at 69.3 h. This period wherein the substrates are co-utilised shortly prior to the complete depletion of the preferred substrate (glucose) can be interpreted as a resource allocation strategy to minimise lost growth during the lag phase (e.g. by investing greater carbon resources into energy intensive catabolite sensors) at the cost of future growth potentials (Chu and Barnes, 2016; Egli et al., 1993). This trait has been associated with microorganisms that have evolved to rapidly changing nutrient composition and high competitive stresses. This could be explained by *F. venenatum* A3/5's natural growth environment of soil, where nutrient and microbial community composition exhibit large spatiotemporal heterogeneity (Bonner et al., 2018; Malik et al., 2019; Salomé et al., 2010). Between 88 and 95.7 hours, the remaining xylose is almost fully depleted leading to a sharp decline in the specific growth rate as the biomass transitions into the stationary phase (95.7 to 121.3 h).

### 3.1.2. *Variability in the time-series data*

Replicate variability differed significantly across the fermentation time course. Glucose and xylose concentrations for the 1:1 condition exhibited the lowest variability of state variables across the time-series, with overall coefficients of variation ($\overline{CV}$) equal to 0.037 and 0.041 respectively, while for the 2:1 condition there was a marked increase in overall variation ($\overline{CV}$ = 0.092 and 0.112 respectively). The variability in substrate concentrations, particularly during the log phases, likely arose from biological stochasticity and the high rate of change of substrate in the environment during this period rather than due to significant analytical error, as the reproducibility of the calibration points was very high for both glucose and xylose standards (data not shown). However, glucose concentration became undetectable below 0.6 g/L due to interference with a closely eluted media compound, suggesting further refinement of the protocol is required to improve signal resolution.

For both conditions, biomass concentrations demonstrated a moderate degree of variability ($\overline{CV}$ = 0.114 and 0.086) across the time-series, particularly for biomass concentrations corresponding to $OD600$ values between 0.657 (min, 59.7 h) and 2.724 (max, 121.3 h). This is potentially a result of diminished resolution of OD600 absorbance measurements for high density cultures, as reflected in the non-linearity of the obtained $CDW$ to $OD600$ calibration (see Supplementary Materials) (E. Lindstrom et al., 1998; Kensy et al., 2009). This limitation is likely compounded by the increasing morphological heterogeneity (in both size and shape) of the hyphal cells post-germination (Stevenson et al., 2016). Furthermore, pellicle formation at the air-liquid interface was visible in all wells of the 96-well plate after 69.3 h (see Supplementary Materials), demonstrating significant spatial biomass heterogeneity in the culture volume.

### 3.2. Bootstrap sample parameter estimation and model performance evaluation

#### 3.2.1. Biokinetic parameter estimates

The parameters of the biokinetic model system were estimated using a bootstrap approach with 1500 samples from the 1:1 glucose-xylose data. Table 2 presents the summary statistics of the parameter estimate distributions, including the mean, median, standard deviation, and the 2.5th and 97.5th percentiles. Fig. 4 illustrates the histograms of the bootstrap parameter estimate distributions, with visual indicators for the mean, median, and confidence intervals, giving a comprehensive view of the variability and central tendency of each estimated parameter.

**Table 2.** Summary statistics of the parameter estimate distributions determined from 1500 bootstrap samples of the original 1:1 glucose-xylose data. Values are presented to 3.s.f.

| Parameter | Units | Mean | Median | Std Dev | 2.5th Percentile | 97.5th Percentile |
|---|---|---|---|---|---|---|
| $\mu_{m1}$ | h$^{-1}$ | 0.113 | 0.115 | 0.00892 | 0.0968 | 0.129 |
| $K_{c1}$ | g·L$^{-1}$ | 0.0851 | 0.0712 | 0.0383 | 0.047 | 0.195 |
| $\mu_{m2}$ | h$^{-1}$ | 0.0362 | 0.0335 | 0.011 | 0.0231 | 0.0691 |
| $K_{c2}$ | g·L$^{-1}$ | 0.235 | 0.144 | 0.241 | 0.0547 | 0.969 |
| $k_I$ | g·L$^{-1}$ | 1.5 | 1.27 | 0.83 | 0.529 | 3.54 |
| $Y_{S1}$ | g (glucose)·g (biomass)$^{-1}$ | 2.54 | 2.54 | 0.173 | 2.22 | 2.87 |
| $Y_{S2}$ | g (xylose)·g (biomass)$^{-1}$ | 4.12 | 4.07 | 0.644 | 3.13 | 5.54 |

The mean maximum specific growth rate for glucose ($\mu_{m1}$) was 0.113 h$^{-1}$ with a narrow confidence interval, while the maximum specific growth rate for xylose ($\mu_{m2}$) was significantly lower at 0.0362 h$^{-1}$ and shows greater variability with a slightly right-skewed unimodal distribution. The half-saturation constant for glucose ($K_{c1}$) has a relatively low mean value (0.0851 g·L$^{-1}$) compared to xylose ($K_{c2}$), which was higher and more variable (mean = 0.235 g·L$^{-1}$), and the distributions of both parameters were highly skewed to the right. The inhibition constant ($k_I$) distribution demonstrated substantial variability and rightward skewness, with the mean (1.5 g·L$^{-1}$) and median (1.27 g·L$^{-1}$) far from the modal peak, indicating sensitivity in the catabolite repression effect of glucose on xylose uptake. The estimated yield coefficient for glucose ($Y_{S1}$) was lower (mean = 2.54 g (glucose)·g (biomass)$^{-1}$) than for xylose ($Y_{S2}$), (mean = 4.12 g (xylose)·g (biomass)$^{-1}$), suggesting a higher growth efficiency when utilising the preferred substrate, glucose. Both yield constant distributions show high variability, with platykurtic profiles of relatively low skewness.

The large discrepancy in values of $\mu_{m1}$ (0.113 h$^{-1}$) and $\mu_{m2}$ (0.0362 h$^{-1}$) explains the sequential utilisation of glucose and xylose observed in the data. This diauxic pattern is commonly observed in batch systems with high substrate concentrations, where the biomass selectively utilises substrates that support highest maximum growth rates (Egli et al., 1993; Monod and Bordet, 1942).

The estimated value of $\mu_{m1}$ (0.113 h⁻¹) was found to be significantly lower than that reported for industrial production of Quorn mycoprotein from glucose feedstock (0.17 - 0.20 h⁻¹) (Moore et al., 2000), suggesting that the intrinsic value of $\mu_{m1}$ was not identifiable from the mixed substrate fermentation data. This is supported by previous studies proving that $\mu_m$ (and $K_C$) are significantly influenced by the historic physiological and environmental states of the culture. Therefore, it cannot be expected that the estimated value of $\mu_{m1}$ would be the same for cultures containing solely glucose (Chen et al., 2018; Ghosh and Pohland, 1972; Peil and Gaudy, 1971).

The relatively low value of $\mu_{m2}$ (and larger values of $Y_{S2}$ and $K_{S2}$) compared to glucose could be attributed to lower biosynthesis efficiency of the different metabolic pathways for secondary substrate utilisation. Mendonca et al. (2024) performed metabolic flux analyses for lignocellulosic glucose and p-coumarate in soil-borne *P. putida* and found significantly lower carbon utilisation efficiency for p-coumarate, with over 70% of p-coumarate carbon converted to $CO_2$ (compared to ~25% of glucose). Furthermore, although not included in our model, trace quantities of extracellular xylitol (<1 g/L, data not shown) were detected in the culture media during the second log phase. It has been found that xylitol accumulation from pentose phosphate pathway conversion of xylose is linked to low xylose transport efficiency, resulting in lower observed growth rates in yeasts (Stambuk et al., 2008; Veras et al., 2019).

The estimated value of $Y_{S1}$ (2.54 g (glucose)·g (biomass)⁻¹) was found to be higher than that reported for Quorn production (0.897 g (glucose)·g (biomass)⁻¹). This is likely due to diversion of glucose resources towards the end of the first log phase towards activation of biosensors and upregulation of transporter proteins/enzymes for the uptake of xylose, thereby reducing the overall yield of biomass from glucose (discussed in Section 3.1.1).

Furthermore, the estimated values of both $Y_{S1}$ and $Y_{S2}$ are similar to those reported for other mycoprotein-producing species in mixed substrate cultures. Specifically, *Pleurotus sp.* cultured on apple, spinach, and beet substrates consumed 4 g substrate per g biomass (Ahlborn et al., 2019), and *Paradendryphiella salina* cultured on seaweed and seaweed waste consumed 1.78 g substrate per g biomass (Landeta-Salgado et al., 2021). The relative values of $Y_{S1}$ and $Y_{S2}$ were also found to be similar for lignocellulosic cellobiose and xylose utilisation by *S. cerevisiae* (Chen et al., 2018). These results indicate that *F. venenatum* A3/5 yields are consistent with those observed for similar organisms utilising waste substrates, and further highlight parameter dependency on culture environment.

The estimated value of $K_{C1}$ (0.0851 g/L) is an order of magnitude higher than was previously reported for *F. venenatum* (0.0054 g/L) by Wiebe et al. (1992), suggesting an underestimation of glucose affinity at low concentrations. However, it should be noted that in the same study, $K_C$ values were found to vary significantly with different competitive stresses (varying dilution rates and limiting nutrients), so it is possible that the presence of xylose in our study may have influenced this parameter. Furthermore, the experimental design was fundamentally different in the previous work which utilised long-term continuous cultivation. Long-term chemostat cultures are more likely to reflect intrinsic $K_C$ values as cells adapt over time to efficiently utilise the limiting substrate, often resulting in lower $K_C$ values. This adaptation improves substrate affinity at the cost of lowering the maximum specific growth rate (Kovárová-Kovar and Egli, 1998; Senn et al., 1994). It is also possible that the poor analytical capacity of the UPLC protocol in detecting small concentrations of glucose may have decreased the accuracy of this parameter value (Shehata and Marr, 1971). Interestingly, the extreme skewness observed in the distributions of $K_{C1}$ and $K_{C2}$ with modal densities close to the lower realistic boundary was also found by Manheim et al. (2019) in modelling microcystin biodegradation in drinking water treatment. This clustering indicates poor identifiability for both half-saturation constants.

Despite the substantial variation in $k_I$ due to the variability in replicate data at the onset of xylose uptake, the mean estimated value appears reasonable for this system. This $k_I$ value (1.5 g.L$^{-1}$) suggests that catabolite repression mechanisms diminish, and xylose uptake begins when glucose concentration falls below 1.5 g/L. This interpretation is consistent with the calibration data, which shows significant xylose consumption starting at glucose concentrations between 1.344 and 1.583 g/L (see Supplementary Materials).

### 3.2.2. Parameter correlation analysis

The parameter correlation analysis indicated several significant relationships between the estimated parameters, which are summarised in heatmap correlation matrix (Fig. 5).

There was a strong, positive correlation between $K_{c1}$ and $k_I$ (r = 0.75, p < 0.001) and between $\mu_{m2}$ and $Y_{S1}$ (r = 0.80, p < 0.001). Additionally, $\mu_{m2}$ and $K_{c2}$ exhibited a strong, positive correlation (r = 0.69, p < 0.001). These correlations suggest a significant interdependence in the efficiency of substrate uptake and metabolic adaptation. Conversely, strong negative correlations were observed between $Y_{S1}$ and $Y_{S2}$ (r = -0.73, p < 0.001) and between $\mu_{m2}$ and $Y_{S2}$ (r = -0.65, p < 0.001). These indicate a trade-off between the yield coefficients for glucose and xylose and between the specific growth rate on xylose and its yield coefficient. Moderate positive correlations included $\mu_{m2}$ and $k_I$ (r = 0.46, p < 0.001), $K_{c2}$ and $k_I$ (r = 0.49, p < 0.001) and $\mu_{m1}$ and $K_{c1}$ (r = 0.42, p < 0.001). Moderate negative correlations were found between $\mu_{m1}$ and $\mu_{m2}$ (r = -0.32, p < 0.001). These relationships highlight further interactions between the specific growth rates, half-saturation constants, and the inhibition constant.

The presence of these strong and moderate correlations suggests that the simultaneous estimation of unique values for multiple parameters could be challenging to achieve with the current experimental set-up. This interdependence therefore must be considered when interpreting the parameter estimation results and in design of future experiments. For example, the correlation between $\mu_m$ and $K_c$ is a prevalent issue in many works, and several design strategies have been proposed to obtain independent estimates (Grady et al., 1996; Healey, 1980; Lobry et al., 1992).

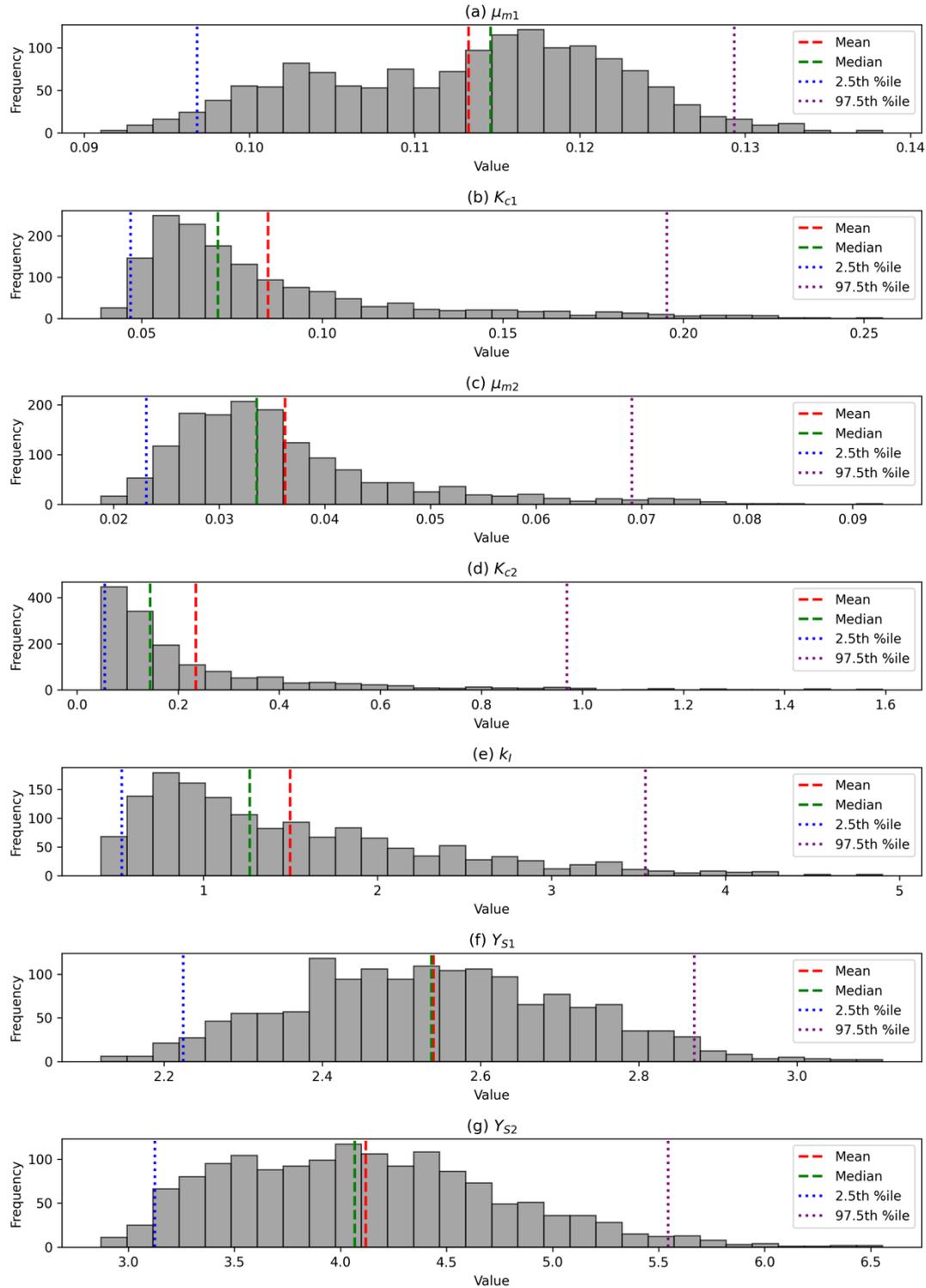

**Fig. 4.** Histograms of the Bootstrap Parameter Estimates Obtained Using 1:1 Glucose-Xylose Experimental Data. Each subplot corresponds to a specific parameter: (a) $\mu_{m1}$ (b) $K_{c1}$ (c) $\mu_{m2}$ (d) $K_{c2}$ (e) $k_I$ (f) $Y_{S1}$ and (g) $Y_{S2}$. The histograms display the distribution of parameter estimates obtained from a total of 1500 bootstrap samples. Summary statistics are represented as: mean (red dashed line), median (green dashed line), 2.5th percentile (blue dotted line), and 97.5th percentile (purple dotted line). (For interpretation of the references to colour in this figure legend, the reader is referred to the Web version of this article).

One such strategy is to increase the ratio $S_0/X_0 > 20$ so the maximum potential of the population to utilise the limiting substrate can be observed. As the ratio was well above this value for the current study ($S_0/X_0 = 150$), this rule clearly does not hold true for mixed substrate cultivations. Indeed, it cannot be said that substrates were utilised to their maximum growth potentials, as energy was expended in the switch from glucose to xylose utilisation (as discussed in Section 3.1.1). Therefore, future research could explore the estimates of $\mu_m$ and $K_c$ values, their estimates should first be obtained from single substrate batch cultivations with high substrate to biomass ratios. The estimates can then be used as fixed priors for the mixed-substrate estimation problem, serving to isolate the intrinsic kinetics of *F. venenatum* in complex media.

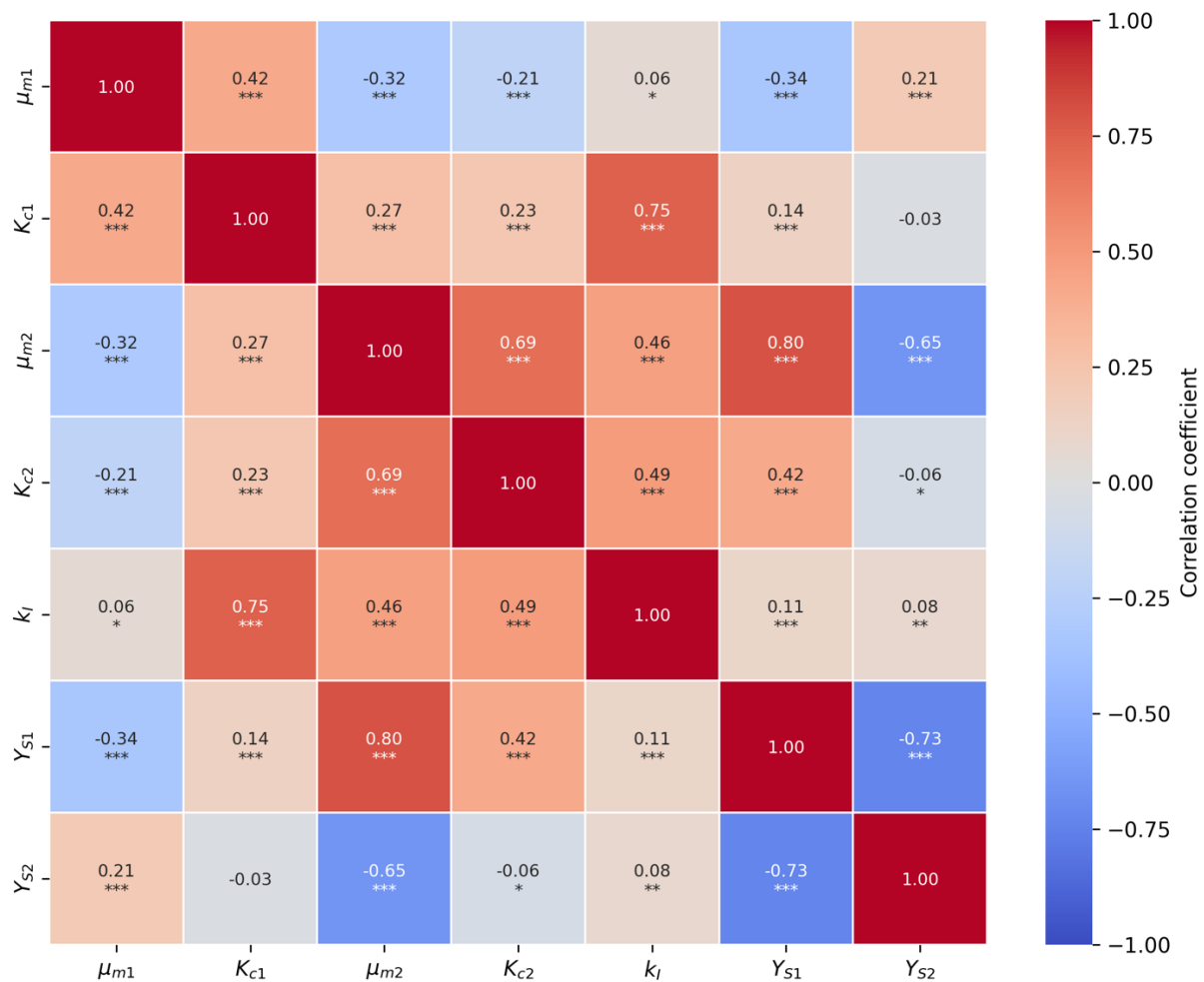

**Fig. 5.** Parameter Correlation Heatmap with Significance Levels. Pearson correlation coefficients were determined for pairwise parameters estimated from the bootstrap sample distributions of the dual substrate fermentation model. Each cell provides the correlation coefficient value and the level of statistical significance: ***p < 0.001, **p < 0.01, *p < 0.05. Cells without asterisks are not significant. The diagonal values are 1.0 by definition.

### 3.2.3. Goodness of fit and predictive performance

The performance metrics of model simulations for both 1:1 and 2:1 glucose-xylose conditions are summarised in Table 3.

**Table 3.** Performance Metrics of Model Simulations for 1:1 and 2:1 Glucose-Xylose Conditions. Performance metrics coefficient of determination ($R^2$), mean absolute error (MAE), and root mean squared error (RMSE) for biomass (X), glucose ($S_1$), and xylose ($S_2$) concentrations are presented. The 1:1 glucose-xylose condition metrics indicate the model performance based on the bootstrap parameter estimate distributions, while the 2:1 glucose-xylose condition metrics validate the model's generalisability to unseen data.

| State variable | 1:1 Glucose-xylose (calibration) | | | 2:1 Glucose-xylose (validation) | | |
|---|---|---|---|---|---|---|
| | $R^2$ | MAE | RMSE | $R^2$ | MAE | RMSE |
| Biomass concentration ($X$) | 0.986 | 0.334 | 0.413 | 0.856 | 0.922 | 1.288 |
| Glucose concentration ($S1$) | 0.986 | 0.415 | 0.640 | 0.941 | 1.272 | 2.016 |
| Xylose concentration ($S2$) | 0.996 | 0.275 | 0.387 | 0.949 | 0.622 | 0.957 |

Fig. 6 presents the simulation of the model system using the parameter estimates obtained from bootstrap sampling plotted against the 1:1 glucose-xylose data used for calibration. The simulation shows a high goodness of fit for each state variable, with $R^2$ values of 0.986 for biomass ($X$), 0.986 for glucose ($S_1$) and 0.996 for xylose ($S_2$). The corresponding RMSE values are 0.413 g/L, 0.640 g/L, and 0.387 g/L, respectively. The model accurately captures the system dynamics critical to fermentation process design, including the log phases. The diauxic shift is distinctly captured, underscoring the model's suitability in describing the underlying catabolite repression dynamics. Model performance was superior to previous work studying yeast astaxanthin production from glucose and sucrose (Vega-Ramon et al., 2021). This is likely due to improved ability of the stochastic global algorithm (DE) used in this work to identify global minima compared with local gradient descent methods, which are known to demonstrate poor exploration of search region and get trapped in local minima for highly non-linear problems (Ashraf and Pfaendtner, 2020; Fang et al., 2009; Sumata et al., 2013). Furthermore, the previous work also modelled astaxanthin production with a further set of parameters increasing the optimisation complexity.

However, the fitness metrics (particularly for $S_2$) suggest potential overfitting, likely arising from a common issue where the optimisation algorithm gives preferential weighting to one variable. This tendency has been well documented in the literature and could be addressed through alternative objective function formulations or optimisation methods and pipelines (Manheim and Detwiler, 2019). The model captures well the regions of variability in the experimental data while retaining the expected dynamic profile of system.

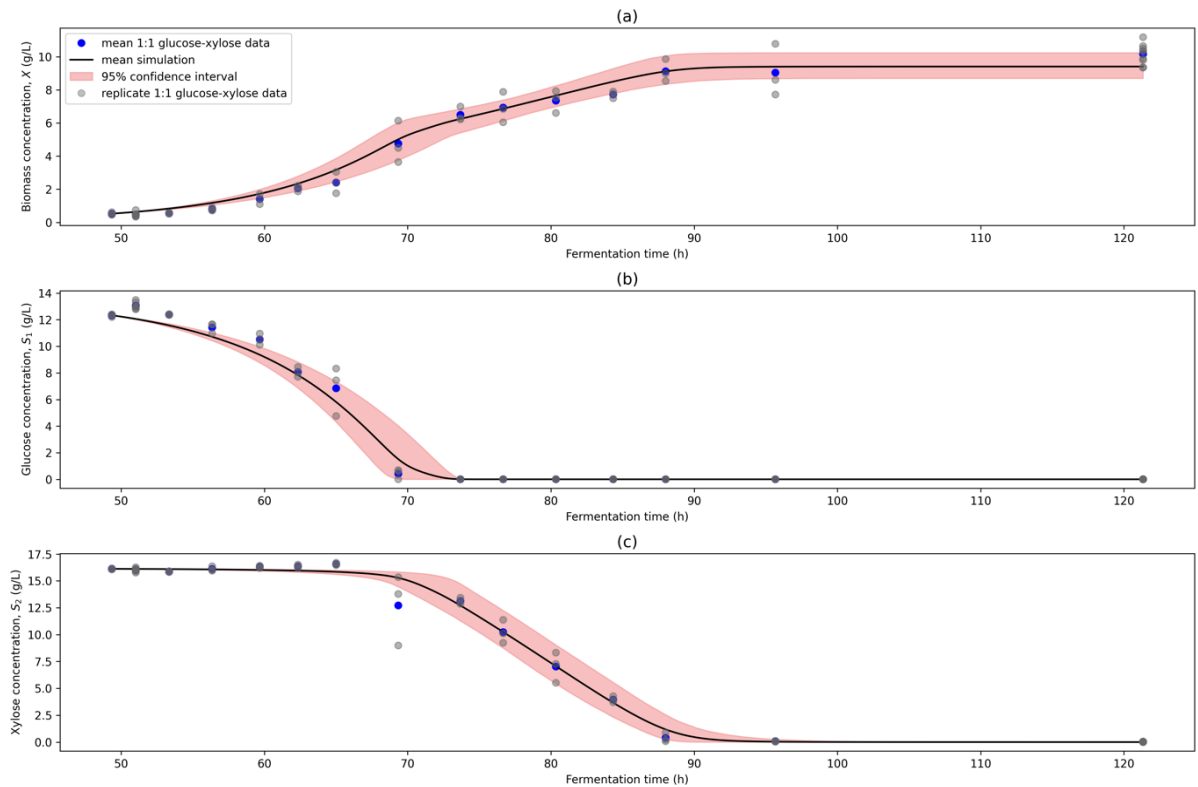

**Fig. 6.** Model Calibration Simulation Plot. (a) biomass ($X$), (b) glucose ($S_1$), and (c) xylose ($S_2$) concentrations simulated over the fermentation time. Mean (black line) and 95% confidence intervals (red band) of model simulations from 1500 bootstrap sample parameter estimates are compared to 1:1 glucose-xylose experimental data (mean data = blue dot; replicate data = grey dot). (For interpretation of the references to colour in this figure legend, the reader is referred to the Web version of this article).

Fig. 7 shows the model simulation using the parameter estimate distributions from bootstrap sampling compared with the 2:1 glucose-xylose data for validation. The results indicate a good model fit to the test condition data for each state variable, with $R^2$ values of 0.856 for biomass, 0.941 for glucose, and 0.949 for xylose, demonstrating good model generalisability. While the model fit is more balanced between variables during the lag phase and early log phases, the effects of overfitting in the calibration step are clearly demonstrated in the second log and stationary phases, where biomass and glucose concentrations are over- and underestimated respectively, The superior fit to $S_2$ compared to $S_1$ and $X$ highlights the greater weight given to this variable during calibration, adversely affecting the model's generalisability to unseen conditions. The high uncertainty and relatively low goodness of fit for biomass prediction are also explained by the broad distributions of the yield coefficients and inhibition constant estimates.

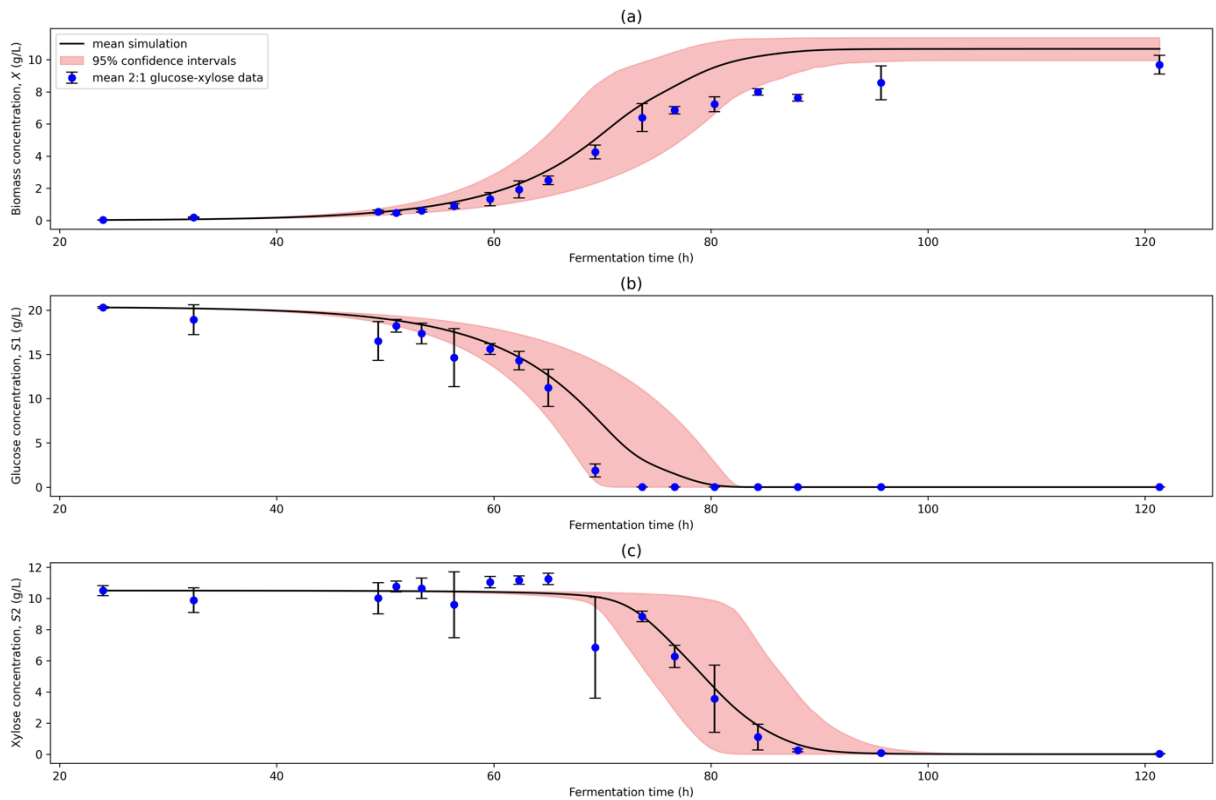

**Fig. 7.** Model Validation Simulation Plot. (a) biomass ($X$), (b) glucose ($S_1$), and (c) xylose ($S_2$) concentrations simulated over the fermentation time course. The initial value problem for validation was solved using $[\overline{X}, \overline{S_1}, \overline{S_2}]_{t=24}$ from the 2:1 glucose-xylose time series dataset. Mean simulation (black line) and 95% confidence intervals (red band) using 1500 bootstrap sample parameter estimates are compared to mean aggregated 2:1 glucose-xylose experimental data (blue dots). (For interpretation of the references to colour in this figure legend, the reader is referred to the Web version of this article).

Overall, the model demonstrates a good capacity to predict the fermentation dynamics of *F. venenatum* in environments containing a mixture of glucose and xylose substrates at varying concentrations.

## 4. Conclusions and future work

This study provides a detailed analysis of the biokinetics of *Fusarium venenatum* A3/5 cultivated with a mixture of glucose and xylose, revealing several key findings. Notably, the estimated parameter values highlighted the strain's efficient glucose utilisation and adaptive capability to switch substrates without a significant lag phase in line with previous research findings. The high $R^2$ values for model simulations against both calibration and validation data sets underscore the model's predictive accuracy and potential utility in fermentation process design. High throughput experiments combined with a bootstrap sampling approach allow for rapid determination of parameter estimates and their uncertainties.

However, there are several limitations of the current study to be addressed in future work. The current model does not account for the effects of mixed substrate utilisation on the protein quality of *F. venenatum*, of critical importance for microbial protein food products. Therefore, future work will aim to model the influence of lignocellulosic substrate environments on protein and amino acid composition.

Furthermore, lignocellulosic hydrolysates derived from agricultural waste resources are compositionally diverse (Rambo et al., 2015), containing low concentrations of other sugars (inc. arabinose, galactose, mannose), organic acids, and furfural. Additionally, the composition of lignocellulosic biomass varies significantly depending on the resource type (Da Silva Perez et al., 2015), and hydrolysate composition is influenced by the efficiency of the pretreatment and hydrolysis processes (Coelho et al., 2024). Consequently, the current model does not fully capture the complexity of a true lignocellulosic resource to mycoprotein bioprocess. These factors collectively underscore the need for future models to incorporate the variability and complexity inherent in lignocellulosic biomass to improve their applicability and accuracy in real-world scenarios utilising agricultural residues as feedstock.

To address the identified overfitting and parameter correlation issues, future studies should investigate a broader range of experimental conditions and incorporate more robust methods into the parameter estimation framework, such as k-fold cross-validation. Additionally, alternative optimisation approaches could be explored such as Bayesian optimisation and Gaussian process (GP) modelling, which have been demonstrated in similar systems to improve parameter identifiability and model predictive capacity (Bradford et al., 2020, 2018; Manheim and Detwiler, 2019; Rogers et al., 2023; Vega-Ramon et al., 2021).

Furthermore, there are several drawbacks of the current analytical techniques employed. The precision of OD600 measurements for high biomass densities could be improved by instead using light scattering or fluorescence time-derivatives as a surrogate (Kensy et al., 2009; Krishnamurthi et al., 2021). The limitation of UHPLC in detecting low concentrations and its use for targeted analyses could be overcome by incorporating complementary methods such as LC-MS and GC-MS, which offer high sensitivity and untargeted chemical analysis. Additionally, cost-effective and high throughput techniques like infrared and Raman vibrational spectroscopy could be employed to further enhance analytical capabilities (Metcalfe et al., 2020; Semeraro et al., 2023). Enhancing the analytical protocol could improve the quality of experimental results and mitigate parameter uncertainty.

Overall, the study lays a solid foundation in understanding the biokinetics of mixed substrate fermentation using *F. venenatum* A3/5. High throughput experiments combined with a bootstrap sampling approach allows for rapid determination of parameter estimates and their uncertainties to inform the design and optimisation of a lignocellulosic waste to mycoprotein bioprocess for more sustainable and efficient protein production.

**CRediT authorship contribution statement**

**Mason Banks:** Conceptualization, Data Curation, Formal Analysis, Investigation, Methodology, Project Administration, Software, Validation, Visualization, Writing – Original Draft, Writing – Review & Editing. **Mark Taylor**: Methodology, Resources, Supervision. **Miao Guo**: Funding Acquisition, Resources, Supervision, Writing – Review & Editing.

**Declaration of competing interest**

The authors declare the following financial interests/personal relationships which may be considered as potential competing interests: Mason Banks reports financial support was provided by EPSRC DTP Industrial CASE Studentship (project reference 2609294) where Marlow Ingredients provided top-up funding. If there are other authors, they declare that they have no known competing financial interests that could have appeared to influence the work reported in this paper.

**Data availability**

All data used in this work can be found in Supplementary Materials. Any additional information will be made available on request.

**Acknowledgements**

The authors would like to express gratitude for the support provided by EPSRC DTP Industrial CASE Studentship (project reference 2609294) and Marlow Ingredients for top-up funding. Mason Banks would like to thank the Nikon Imaging Centre at King's College London for providing training and access to the confocal laser scanning microscope used to capture the image shown in Fig. 1. Mason Banks would like to thank Louisa Edge and Nicola Thomas (Fermentation Team, Marlow Ingredients) for training in experimental methods and techniques. Mason Banks and Miao Guo would like to thank Anastasiia Zaleska (Photonics & Nanotechnology Group, Department of Physics, King's College London) for providing support and access to the light microscope used for conidia counting.